# Systematic Review Protocol

# Requirements Engineering in Quantum Computing

Final version:  20/11/2023

Dr. Samuel Sepúlveda, Dra. Ania Cravero
Dpto. Cs. de la Computación e Informática
Facultad de Ingeniería y Ciencias
Universidad de La Frontera
Temuco, Chile.

**Abstract**

*Context*: Quantum computing (QC) represents a paradigm shift in computational capabilities, presenting unique challenges in requirements engineering (RE). The complexity of quantum systems and rapid technological advancements necessitate a comprehensive understanding of the current state and future trajectories in RE for QC.

*Objective*: A protocol for carrying out a systematic literature review about the evidence for identifying and analyzing the challenges in RE for QC software. It seeks to evaluate the current methodologies employed in this domain and propose a forward-looking perspective on the evolution of these methodologies to meet future industry and academic needs.

*Method*: This protocol employs a structured approach to search and analyze relevant literature systematically, according to Barbara Kitchenham's guidelines.

*Results*: A validated protocol to conduct a systematic review. The protocol is expected to yield diverse literature spanning theoretical frameworks, empirical studies, and methodological advancements in RE for QC. It will highlight the current challenges, opportunities, and future directions, offering insights into the field's academic and practical aspects.

*Conclusions*: The systematic review aims to provide a nuanced understanding of the RE landscape in QC. It will offer valuable insights for academic researchers, industry professionals, software engineers, industry analysts, and educators, shaping the future discourse in QC development.

# 1.    Introduction

Today, we stand at the cusp of a new paradigm in computational development – Quantum Computing (QC). This paradigm, grounded in quantum mechanics principles, promises computational capabilities far beyond classical computers' reach (Courtland, 2017). The unique nature of QC requires new models and methodologies in Software Engineering (SE) to address its distinct characteristics and challenges (Serrano, Perez-Castillo, et al., 2022).

The need for new development models and methodologies in Quantum Software Engineering (QSE) is justified by QC's unique challenges compared to classical computing (Piattini, Serrano, et al., 2021). These specialized approaches are crucial for addressing quantum algorithms' complexity, the probabilistic nature of qubits, scalability, optimization, and integration with existing technologies.

Despite QC's rapid evolution and transformative potential, a significant gap exists in tools and methodologies for developing software to leverage this potential fully. This gap has spurred the necessity to research and propose solutions for a smoother transition between classical and quantum paradigms. The primary goal is to bridge traditional software development methodologies with QC's peculiarities, allowing for a more effective and coherent transition (Felderer, Taibi et al., 2023).

To address the increasing demands and requirements of the software industry, Software Engineering (SE) plays a crucial role in overseeing all aspects of software product and service development. Specifically, Requirements Engineering (RE) focuses on the identification, modeling, communication, and documentation of system requirements and the context of their usage, as highlighted by (Paetsch, Eberlein, et al., 2003). RE is defined as the SE branch that addresses how to meet "real world" objectives through software system functions and constraints, a concept articulated by (Zave, 1997). This field faces numerous challenges, including incomplete or hidden requirements and fluctuating needs, as noted by (Fernández, Wagner, et al., 2017; Ambreen, Ikram, et al. 2018). An initial identification and discussion about the new competencies required for effective RE practices in complex systems is presented by (Jantunen, Dumdum, et al., 2019). This work states how increased complexity impacts RE practices and the identification of critical viewpoints essential for aligning RE practices with specific design problems.

Recognizing the challenges quantum software developers face is crucial to shaping this field. Traditional software development methodologies may not seamlessly transition to the quantum realm, necessitating specific approaches for QC that consider phenomena like superposition and entanglement. Quantum programming remains complex, and new algorithm formulation is intricate. Reliable quantum systems demand formal techniques in software development, providing precise bases for algorithm development and analysis (De Stefano, Pecorelli et al., 2022). Although progress is being made in quantum software tools, a comprehensive toolchain for quantum software development is still in its infancy (Alexeev, McCaskey et al., 2022; Alexeev, McCaskey et al., 2023). Thus, Quantum Software development requires tailored methodologies and tools focused on addressing specific challenges of quantum software and ensuring system reliability.

This work is part of the research project GI23-0012, aimed at disseminating and promoting quantum computing and informatics. The project's primary goal is to integrate these crucial concepts into the educational curriculum of future software developers. This initiative is spearheaded by academics from the Department of Computer Science and Informatics at the Universidad de La Frontera. Through this project, the aim is to equip students with the necessary tools and knowledge to meet the emerging challenges in the field of quantum computing, preparing them to innovate and lead in this rapidly evolving area.

The rest of this report is structured as follows. Section 2 describes the details of the protocol definition for carrying out the systematic review. Finally, Section 3 presents the conclusions and future work.

## 2.    Protocol definition

This section describes the protocol definition of a systematic review (SR). This protocol considers the three main phases (planning, conducting, and reporting), according to the protocol defined by Bárbara Kitchenham  (Kitchenham, 2004; Kitchenham and Charters, 2007; Kitchenham and Madeyski, 2022).

### 2.1.    Planning

The planning phase consists of declaring the aim, need and research questions (RQs) for the SR. Also, we define the search string to gather the relevant papers for reviewing.

#### 2.1.1.    Aim and need

The SR aims to identify and analyze the unique challenges inherent in RE for QC software, evaluate the current methodologies being employed to address these challenges and propose a forward-looking perspective on the evolution of these methodologies to meet future needs in the field.

The expected impact of the SR is to provide a comprehensive overview of the state-of-the-art RE practices tailored to QC, highlight the gaps between traditional and quantum software development, and offer a roadmap for researchers and practitioners to navigate the complexities of this emerging paradigm.

The motivation for this systematic review stems from the transformative nature of QC, which introduces complex challenges in RE that are distinct from traditional computing. This review seeks to meticulously analyze these challenges, evaluate existing methodologies, and propose future directions. Its impact is expected to guide the evolution of RE practices to meet quantum computing demands effectively. This effort will substantially contribute to theoretical understanding and practical advancements in this rapidly evolving field.

#### 2.1.2.    Research questions

The context for the research questions is centered around the emergent field of quantum computing. This new computational paradigm offers capabilities that surpass those of classical computing systems. Quantum computing enables parallel processing and significantly enhanced computational speed. This technological leap brings forth unique challenges in requirements engineering, necessitating new models and methodologies distinct from those used in classical software engineering. The research questions are formulated to investigate these specific challenges, explore the current state of methodologies in quantum computing, and identify future directions for requirements engineering within this novel and complex domain.

The research questions arise from the main RQ: What are the challenges, opportunities, and future directions on requirements engineering for quantum computing software? This RQ is divided into three more detailed RQs, as shown in Table 1.

**Table 1.** RQs and aim.

| RQ# | Research Question | Aim | Possible answers or classification schema |
|---|---|---|---|
| RQ1 | What specific challenges are currently faced in RE for QC? | To identify and understand the unique challenges that QC poses to RE. | <ul><li>Technical Complexity</li><li>Evolution and Scalability</li><li>Integration Requirements</li><li>Human Resource Constraints</li></ul> |
| RQ2 | What opportunities does QC present for advancements in RE? | To explore how current RE methodologies are being adapted to suit the needs of QC. | <ul><li>Methodology Types (Agile, Waterfall, etc.)</li><li>Tooling and Technologies</li><li>Stakeholders Involvement</li></ul> |
| RQ3 | What future directions or trends are emerging in the field of RE specific to QC? | To identify and forecast potential innovations and future directions in RE specifically tailored for QC. | <ul><li>Model-Based Innovations (QC Requirement Models, Simulation and Visualization Tools)</li><li>Methodological Innovations (Hybrid Q-Classical RE Methodologies, Adaptive and Iterative RE Frameworks)</li><li>Tool and Technique Innovations (Requirement Visualization for QC, Automated Tools for Requirement Analysis)</li><li>Collaboration and Communication Innovations (Interdisciplinary Collaboration Models, Stakeholder Engagement Strategies)</li><li>Education and Training Innovations (RE Training Programs for QC, Educational Resources and Workshops)</li></ul> |

### 2.1.3. Search string

According the Kitchenham's guidelines, we consider:

- extracting keywords from the context and RQs
- considering synonyms for the keywords
- using the PICOC criterion to build the search string (Petticrew and Roberts, 2008; Zhang, Babar et al., 2011). See the details in Table 2.

From the RQs, we derive keywords: "quantum computing," "requirements engineering," "challenges," and "future directions."

Then, consider synonyms and related terms for these keywords to widen the search scope.

**Table 2.** PICOC criterion, details and application.

| Criteria | Scope | Detail in SE | Application to our case |
|----------|-------|--------------|-------------------------|
| Population | Who?/What? | For SE should correspond to one of the following: (1) specific SE role, (2) a category of software engineer, (3) an application area or (4) an industry group. | QC software projects |
| Intervention | How? | In SE is defined as a methodology, tool, technology or procedure that addresses a specific issue. | RE methodologies and approaches |
| Comparison | Compare to…? | N/A | Not applicable, the study doesn't compare interventions. |
| Outcomes | What …to accomplish? effect? | Outcomes should relate to factors of importance to practitioners such as improved reliability, reduced production costs, and reduced time to market. | Identified challenges, opportunities, and directions in RE for QC. |
| Context | Under what circumstances? | For SE, this is the context in which the comparison takes place, the participants taking part in the study, and the tasks being performed. | The application in both academic research and industry practice |

An initial search string could be:

```
("quantum computing" OR "quantum software") AND
"requirements engineering" AND ("challenges" OR
"opportunities" OR "future directions")
```

### 2.1.4. Protocol validation

Initially, one of the researchers built the protocol. Then, the rest of the researchers must evaluate and discuss the correctness and completeness of the protocol. Next, the researchers will agree on the changes and corrections, and a validated protocol version will be compiled.

According to the guidelines, we pretend to evaluate the consistency of the protocol. To do this, we have to answer yes to these questions:

- Are the search strings appropriately derived from the RQs?
- Will the data to be adequately extracted address the RQs?
- Is the data analysis procedure appropriate to answer the RQs?

Ideally, and if time permits, we intend to send the improved protocol version to an external expert for review.

## 2.2. Conducting

This phase considers defining the search strategy, the inclusion/exclusion criteria, and the data extraction process. Also, it considers a quality assessment of the gathered evidence and an initial analysis of the threats to the validity of the results.

### 2.2.1. Search strategy

The following sources, detailed in Table 3, are recognized among the most relevant in the SE research community.

**Table 3.** Data sources.

| Source | Link |
|---|---|
| ACM Digital Library | https://dl.acm.org/ |
| IEEE Xplore | https://ieeexplore.ieee.org/Xplore/home.jsp |
| Science Direct | https://www.sciencedirect.com/ |
| Springer Link | https://www.springer.com/ |

According to the recency of the subject and initial searches, the search considers 2018-2023.

The selected language for the search is English, considering that the trending journals and conferences publish their works in this language.

We consider a search process consisting of two phases:
1. Automatic search applying the search string on selected data sources, considering:
   a. different versions of the same paper, we keep the last one
   b. eliminate duplicates
2. snowballing process, using the list of selected papers as seed.

The remaining papers go through the screening phase; the researchers must decide if it is relevant for the SLR. We will review the title, abstract, and keywords to do this. The papers superating this phase will be filtered using the exclusion criteria (EC). For details of the EC, see Table 4.

**Table 4.** Exclusion criteria (EC).

| EC | Description |
|---|---|
| EC1 | The paper is not written in English. |
| EC2 | The paper is not peer reviewed (posters, tutorials, slides, PhD or master thesis and any piece of work considered as grey literature). |
| EC3 | The paper is a secondary study (eventually considered in the related work section). |
| EC4 | The paper is a short paper (four or less pages). |

| EC5 | The focus of the paper is not on proposals treating RE on QC. |
|------|------|

### 2.2.2.  Data extraction

For the final list of selected papers, the following data will be extracted.

- Meta-data for each paper: title, author, publication year, type of publication (journal or conference), results, and future work.
- Detailed-data to answer the RQs. The details are presented in Table 5.

**Table 5.** RQs - Data extraction form.

| Paper # | RQ1 | Evidence RQ1 | RQ2 | Evidence RQ2 | RQ3 | Evidence RQ3 |
|---------|-----|--------------|-----|--------------|-----|--------------|
| 01 | value | | value | | value | |
| 02 | value | | value | | value | |
| … | … | | | | … | |

### 2.2.3.  Graphical summary

To consolidate and ease the data visualization, we consider using a set of tools, including a weighted cloud tag, a Sankey diagram that represents the relationships that existed between the RQs, a map showing the relationship between the main concepts and authors, the most relevant concepts and their relations, among others. The mentioned graphical support considers using tools like VOS-viewer, Sankeymatic, Termine, etc.

### 2.2.4.  Quality assessment (QA)

In order to address bias, each selected paper will be subjected to a QA. To answer about the QA, we will use the criteria proposed by (Dyba and Dingsoyr 2008). This criterion is summarized in five questions that can be answered with yes/partially/no. Other papers have dopted the same criteria (Sepúlveda, Cravero et al. 2015). The QA is presented in Table 6.

**Table 6.** Quality Assessment questions.

| QA# | QA question | Yes ## (%) | No ## (%) | Partially ## (%) |
|-----|-------------|------------|-----------|------------------|
| QA1 | Is the aim of the research sufficiently explained? | … | … | … |
| QA2 | Is the paper based on research methodology? | … | … | … |
| QA3 | Is there an adequate description of the context in which the research was carried out? | … | … | … |
| QA4 | Are threats to validity taken into consideration? | … | … | … |
| QA5 | Is there a clear statement of findings? | … | … | … |

### 2.2.5.  Threats to validity

We consider the types of validity according to (Petersen, Gencel, et al., 2013): theoretical, descriptive, interpretative, generalizability, and reliability.

As a complementary quality indicator, we consider using the PRISMA Statement. This checklist will validate the structure and sections for our SLR report.

### 2.3.  Reporting

To disseminate the results of this work, the results reporting stage considers two steps.
1.  Publish the revised version of the SLR protocol on the arXiv platform.
2.  Publish the SLR results in a high-impact journal, ideally JCR-WoS.

According to the guidelines for conducting and reporting an SLR, the main sections considered are:

1) Introduction
    a) context
    b) motivation
    c) aim and need
    d) structure of the paper.
2) Background
3) Related work: summary of the main related work and RQs answered.
4) Methodology: explain step by step how the protocol will be carried out.
5) Results and discussion
    a) main findings and results
    b) answers to RQs and PQs
    c) QA analysis
    d) threats validity to the paper
6) Conclusions
    a) conclusions
    b) further lines of research.
7) Acknowledgements, recognizing the support for the paper.
8) References
    a) list of bibliographic references
    b) list of selected papers.

## 3. Conclusions and future work

We presented a proposal for the protocol definition of an SLR to summarize and synthesize the evidence about the unique challenges, opportunities, and future directions inherent in RE for QC software.

We pretend to synthesize the current methodologies being employed to address these challenges and propose a forward-looking perspective on the evolution of these methodologies to meet future needs in the field.

## Acknowledgments

The authors thank to research project GI23-0012 supported by Vicerrectoría de Investigación y Postgrado, Universidad de La Frontera.